\documentclass[aps,twocolumn,showpacs,superscriptaddress,floatfix]{revtex4}
\usepackage{graphicx,amsmath,amssymb,times}
\usepackage{txfonts}
\usepackage{psfrag}

\begin{document}
    \newcommand{\p}[1]{\textrm{#1}}
  
  \title{Classification of zero-energy resonances by dissociation of Fesh\-bach molecules}

  \author{Thomas M. Hanna}
  \affiliation{Clarendon Laboratory, Department of Physics,
    University of Oxford, Parks Road, Oxford, OX1 3PU, UK}
     \author{Krzysztof G\'oral}
  \affiliation{Clarendon Laboratory, Department of Physics,
    University of Oxford, Parks Road, Oxford, OX1 3PU, UK} 
   \author{Emilia Witkowska}
  \affiliation{Clarendon Laboratory, Department of Physics,
    University of Oxford, Parks Road, Oxford, OX1 3PU, UK}
  \affiliation{Institute of  Physics, Polish Academy of Sciences, Aleja Lotnik\'ow 32/46, 02-668 
    Warsaw, Poland}   
  \author{Thorsten K\"{o}hler}
  \affiliation{Clarendon Laboratory, Department of Physics,
    University of Oxford, Parks Road, Oxford, OX1 3PU, UK}

  \begin{abstract}
We study the dissociation of Fesh\-bach molecules by a magnetic field sweep across a zero-energy resonance. In the limit of an instantaneous magnetic field change, the distribution of atomic kinetic energy can have a peak indicating dominance of the molecular closed-channel spin configuration over the entrance channel. The extent of this dominance influences physical properties such as stability with respect to collisions, and so the readily measurable presence or absence of the corresponding peak provides a practical method of classifying zero-energy resonances. Currently achievable ramp speeds, e.g. those demonstrated by D\"urr~\textit{et al.} [Phys. Rev. A \textbf{70}, 031601 (2005)], are fast enough to provide magnetic field changes that may be interpreted as instantaneous. We study the transition from sudden magnetic field changes to asymptotically wide, linear ramps. In the latter limit, the predicted form of the atomic kinetic energy distribution is independent of the specific implementation of the two-body physics, provided that the near-resonant scattering properties are properly accounted for. 
    \end{abstract}

  \date{\today}
\maketitle

\section{Introduction}
Cold molecules have been the subject of intense study over the last few years~\cite{review, doyle}. In particular, the manipulation of a magnetic field around a zero-energy resonance has been used for producing both thermal and condensed molecular gases from cold atoms. Typically, the magnetic field is linearly ramped across the resonance~\cite{regal03, str03, cub03, xu03, herbig03, gre03, duerr04prl, MukaiyamaPRL04, DuerrPRA04, volz05}, although other techniques have been demonstrated ~\cite{donley02, chin03, jochim03, jochim03b, mark05, thompson05}.  Detection of cold molecules often relies upon their dissociation, followed by measurement of the resulting atomic gas. Dissociation is commonly performed by photodissociation~\cite{regal03, gre03, chin03} or linearly ramping the magnetic field across the resonance~\cite{regal03, str03, jochim03, jochim03b, cub03, xu03, herbig03, MukaiyamaPRL04, duerr04prl, DuerrPRA04, mark05}. Linear ramps for molecule detection are often slow in order to minimise the speed of the fragments to be imaged. Such sweeps have also been used to determine resonance widths~\cite{DuerrPRA04}. Fast sweeps have been used to measure the population of more than one partial wave~\cite{volz05, duerr05partial}.

In this paper we study in detail dissociation by linear ramps across zero-energy resonances over the whole range of possible ramp speeds. We show that the dissociation process can be used to probe the nature of the resonance. In the limit of a sudden change in the magnetic field strength the distribution of atomic kinetic energy, referred to as the dissociation spectrum, can have a sharp peak at an energy dependent on the final magnetic field~\cite{DuerrPRA04}. Our studies show that such spectra occur only for resonances with weak interchannel coupling. Consequently, resonances may be classified by measuring the peak, which can be done while little is known about the resonance other than its position. This classification scheme applies to Fesh\-bach molecules formed from both Bose and Fermi gases. The strength of the interchannel coupling affects the suitability of the resonance for studying phenomena of current experimental interest. For example, halo molecules such as long-range dimers and three-body Thomas-Efimov states are more easily studied using strongly coupled, entrance-channel dominated resonances~\cite{donley02, koehler03, thomas, efimov, kraemer06, stoll05}, whereas Fesh\-bach molecules created with closed-channel dominated resonances are usually unstable with respect to collisions~\cite{smirne06, MukaiyamaPRL04}. We study the transition from sudden magnetic field changes, referred to as jumps, to asymptotically wide, linear ramps, as the ramp speed and initial and final fields are changed. We provide an analytic treatment of the time evolution for arbitrary ramp speeds and initial and final fields. Taking the limit of an asymptotic ramp produces a spectrum which has been studied both experimentally~\cite{MukaiyamaPRL04, DuerrPRA04} and theoretically~\cite{MukaiyamaPRL04, GKGTJ04, haq05, brouard05}. We show that two-channel and single-channel approaches provide the same asymptotic spectrum. 

In Sections~\ref{sec:2channelapproach} and \ref{sec:spectrum} we introduce the two-channel approach to zero-energy resonances and its application to dissociation spectra, respectively. Dissociation spectra produced by magnetic field jumps and their application to classifying resonances are discussed in Section~\ref{sec:jump}. Section~\ref{sec:asymptotic} is concerned with the transition from jumps to asymptotic ramps. We discuss the different theoretical techniques that may be applied, and the validity of approximating a given, finite ramp to be asymptotic. We conclude in Section~\ref{sec:conclusion}. In the appendices we present the determinations of the time evolution in the two-channel (Appendix~\ref{app:EvolutionOperator}) and single-channel (Appendix~\ref{app:SingleChannel}) approaches. 


\section{Two-channel single-resonance approach}
\label{sec:2channelapproach}

Atom-atom scattering in a cold gas can be magnetically tuned because of the coupling that exists between scattering channels with different spin configurations. The diatomic spin configuration corresponding to the Zeeman state of each atom in the gas is referred to as the entrance channel. This is coupled to the closed channel, so called because its dissociation threshold is far above the relative kinetic energy of atom pairs. The coupled-channels character is reflected in the general form of the two-channel, two-body Hamiltonian of the relative motion,
\begin{align}
H_\mathrm{2B}(B)&=|\mathrm{bg}\rangle H_\mathrm{bg}\langle\mathrm{bg}|
+W|\mathrm{bg}\rangle\langle\mathrm{cl}|\nonumber\\
&+|\mathrm{cl}\rangle\langle\mathrm{bg}|W
+|\mathrm{cl}\rangle H_\mathrm{cl}(B)
\langle\mathrm{cl}|\, .
\label{H2ofBgeneral}
\end{align}
Here ``bg'' and ``cl'' indicate the entrance-channel and closed-channel spin configurations, respectively. $H_\mathrm{bg}$ and $H_\mathrm{cl}(B)$ are the Hamiltonians of the two spin configurations in the hypothetical absence of the interchannel coupling $W$. We choose the zero of energy to coincide at each magnetic field $B$ with the dissociation threshold of the entrance channel. Due to a difference in magnetic moment between the two channels, the closed-channel dissociation threshold can be tuned with respect to zero energy. Consequently, $H_\mathrm{cl}(B)$ contains all of the magnetic-field dependence of $H_\mathrm{2B}(B)$. 

Within the magnetic field range of experimental relevance, typically only one closed-channel state, referred to as the Fesh\-bach resonance state $\phi_{\p{res}}$, has an energy that can become near resonant with pairs of atoms in the entrance channel. Consequently, the closed-channel Hamiltonian $H_\mathrm{cl}(B)$ is well described by taking into account only $\phi_{\p{res}}$, which fulfils the Schr\"odinger equation 
\begin{align}
H_\mathrm{cl}(B) \phi_\mathrm{res} = E_\mathrm{res}(B)
\phi_\mathrm{res} \, .
\end{align}
Here, $E_\mathrm{res}(B) = \mu_{\p{res}}(B - B_{\p{res}})$ is the energy of the Fesh\-bach resonance state. The difference in magnetic moment between the two channels is denoted $\mu_{\p{res}}$, and $B_{\p{res}}$ is the magnetic field strength where the Fesh\-bach resonance state crosses the entrance-channel dissociation threshold, i.e. $E_{\p{res}}(B_{\p{res}}) = 0$. The single-resonance approach to $H_\mathrm{cl}(B)$ consists in the approximation
\begin{align}
H_\mathrm{cl}(B) = |\phi_\mathrm{res}\rangle
E_\mathrm{res}(B) \langle \phi_\mathrm{res}|\,.
\label{eq:hcl}
\end{align}
The two-channel Hamiltonian of Eq.~(\ref{H2ofBgeneral}) then becomes
\begin{align}
H_\mathrm{2B}(B)&=|\mathrm{bg}\rangle H_\mathrm{bg}\langle\mathrm{bg}|
+W|\phi_\mathrm{res},\mathrm{bg}\rangle\langle\phi_\mathrm{res},\mathrm{cl}|\nonumber\\
&+|\phi_\mathrm{res},\mathrm{cl}\rangle\langle\phi_\mathrm{res},\mathrm{bg}|W
+|\phi_\mathrm{res},\mathrm{cl}\rangle E_\mathrm{res}(B)
\langle\phi_\mathrm{res},\mathrm{cl}|\, .
\label{H2BofB}
\end{align} 

A zero-crossing of $E_{\p{res}}$ is accompanied by a singularity in the scattering length $a$ referred to as a zero-energy resonance, at magnetic field $B_0$. This is shifted from $B_{\p{res}}$ by the coupling $W$. On the side of $B_0$ where $a > 0$, the highest excited vibrational bound state $\phi_\mathrm{b}(B)$ supported by $H_\mathrm{2B}(B)$ is usually referred to as the Fesh\-bach molecule, and satisfies
\begin{align}
H_\mathrm{2B}(B) \phi_\mathrm{b}(B)=E_\mathrm{b}(B) \phi_\mathrm{b}(B)
\, . \label{Feshbachmolecule}
\end{align}
Here $E_\mathrm{b}(B)$ is the magnetic-field dependent binding
energy, which vanishes at $B_0$. In the single-resonance approach, the Fesh\-bach molecule assumes the two-component form~\cite{GKGTJ04}
\begin{align}
  \left(
  \begin{array}{c}
    \phi_\mathrm{b}^\mathrm{bg}(B)\\
    \phi_\mathrm{b}^\mathrm{cl}(B)
  \end{array}
  \right)=\frac{1}{\mathcal{N}_\mathrm{b}(B)}
  \left(
  \begin{array}{c}
    G_\mathrm{bg}(E_\mathrm{b}(B))W\phi_\mathrm{res}\\
    \phi_\mathrm{res}
  \end{array}
  \right) \, .
  \label{phib}
\end{align}
Here $\phi_\mathrm{b}^\mathrm{bg}(B)$ and
$\phi_\mathrm{b}^\mathrm{cl}(B)$ are the entrance-channel and closed-channel
components of $\phi_{\p{b}}(B)$, respectively, and the normalisation constant $\mathcal{N}_\mathrm{b}(B)$ is given by
\begin{align}
  \mathcal{N}_\mathrm{b}(B)=\sqrt{1+
    \langle\phi_\mathrm{res}|W
    G_\mathrm{bg}^{2}(E_\mathrm{b}(B))W|\phi_\mathrm{res}\rangle}\, .
  \label{twochannelnormalisation}
\end{align}
In Eqs.~(\ref{phib}) and (\ref{twochannelnormalisation}) the
energy dependent Green's function $G_\mathrm{bg}(z)$,
associated with the entrance-channel Hamiltonian $H_\mathrm{bg}$, is
evaluated at the binding energy $z=E_\mathrm{b}(B)$.
The solution of Eq.~(\ref{Feshbachmolecule}) using the single-resonance Hamiltonian of Eq.~(\ref{H2BofB}) indicates that $E_\mathrm{b}(B)$ fulfils
the condition
\begin{align}
  E_\mathrm{b}(B)=E_\mathrm{res}(B)+
  \langle\phi_\mathrm{res}|WG_\mathrm{bg}(E_\mathrm{b}(B))W|
  \phi_\mathrm{res}\rangle.
  \label{determinationEb}
\end{align}

The continuum part of the spectrum of $H_\mathrm{2B}(B)$ is represented by the scattering states $\phi_\mathbf{p}(B)$, which are labelled by the relative momentum $\mathbf{p}$ of an atom pair and obey
\begin{align}
H_\mathrm{2B}(B) \phi_\mathbf{p}(B)=E \, \phi_\mathbf{p}(B)
\, ,
\label{scatteringstates}
\end{align}
where $E = p^{2}/m$, and $m$ is the atomic mass. The scattering states have
the following two-component form in the single-resonance approach~\cite{GKGTJ04}:
\begin{align}
  \left(
  \begin{array}{c}
    \phi_\mathbf{p}^\p{bg}(B)\\
    \phi_\mathbf{p}^\p{cl}(B)
  \end{array}
  \right)=
  \left(
  \begin{array}{c}
    \phi_\mathbf{p}^{(+)} +
    A(B,E)G_\p{bg}(E+\p{i}0)W\phi_\p{res}\\
    A(B,E)\phi_\mathrm{res}
  \end{array}
  \right) \, .
  \label{phip}
\end{align}
Here, $E+\p{i}0$ indicates that the positive physical energy $E$ is approached from the upper
half of the complex plane, and $\phi_\mathbf{p}^{(+)}$ is the background scattering state of relative momentum $\mathbf{p}$, 
associated with the entrance-channel Hamiltonian $H_\mathrm{bg}$ via \mbox{$H_\mathrm{bg}\phi_\mathbf{p}^{(+)}=E \phi_\mathbf{p}^{(+)}$.} The magnetic-field and energy dependent
amplitude $A(B,E)$ is given by
\begin{align}
A(B,E)=\frac{\langle\phi_\mathrm{res}|W|\phi_\mathbf{p}^{(+)}\rangle}
  {E-E_\mathrm{res}(B)-[\Delta(E)-\mathrm{i}\gamma(E)]}\, ,
    \label{amplitude}
  \end{align}
where the real functions $\Delta(E)$ and $\gamma(E)$ satisfy
\begin{equation}
\Delta(E)-\mathrm{i}\gamma(E) = \langle\phi_\mathrm{res}|WG_\mathrm{bg}(E+\mathrm{i}0)W|
  \phi_\mathrm{res}\rangle\, .
\end{equation}
The function $\Delta(E)$ acquires the meaning of an energy shift, while $\gamma(E)$ captures the decay width of the metastable Fesh\-bach resonance state. In the case of negative energies, as in Eq.~(\ref{determinationEb}), the decay width $\gamma(E)$ vanishes.  


\section{Dissociation spectra}
\label{sec:spectrum}

Fesh\-bach molecules can be dissociated by changing the magnetic field strength to the side of the zero-energy resonance where the Fesh\-bach molecular state does not exist. We consider linear magnetic-field ramps of the form
\begin{equation}
B(t)=B_\p{res}+\dot{B}\,(t-t_\p{res})\, .
\label{eq:bramp}
\end{equation} 
Here $\dot{B}$ is the ramp speed, and $t_{\p{res}}$ is the time satisfying $B(t_{\p{res}}) = B_{\p{res}}$. The ramp begins at the time $t_{\p{i}}$ and ends at $t_{\p{f}}$, ranging between the magnetic fields $B(t_{\p{i}}) = B_{\p{i}}$ and $B(t_{\p{f}}) = B_{\p{f}}$. The ramp of Eq.~(\ref{eq:bramp}) corresponds to a linear time dependence of the resonance energy
\begin{align}
E_\p{res}(t)=\dot{E}_\p{res}\,(t-t_\p{res})\, ,
\label{Eresoft}
\end{align}
where $\dot{E}_\mathrm{res}=\mu_\mathrm{res}\dot{B}$. The validity of
Eq.~(\ref{Eresoft}) presupposes that the electronic degrees of freedom of the atoms adiabatically follow the changing magnetic field strength throughout the dissociation process. The effect of the ramp on the initial molecular state is described by the evolution operator $U_\p{2B}(t_{\p{f}},t_{\p{i}})$, associated with $H_\p{2B}(B(t))$ by the Schr\"odinger equation
\begin{align}
  i\hbar \frac{\partial}{\partial t}U_\p{2B}(t,t')
  =H_\p{2B}(B(t))U_\p{2B}(t,t') \, ,
  \label{eq:U}
\end{align}
with $U_\p{2B}(t,t)=1$. 

In typical experiments~\cite{MukaiyamaPRL04, DuerrPRA04} the atomic gas produced by the sweep is then allowed to freely expand with the magnetic-field strength held at $B_{\p{f}}$, after which a measurement of the atomic momentum distribution is performed. It is therefore instructive to analyse the diatomic state produced by the ramp in terms of the scattering states of Eq.~(\ref{scatteringstates}) evaluated at $B_\p{f}$, which then become stationary states. The transitions into these states due to the ramp determine the dissociation spectrum $n(E)$, in accordance with the general formula~\cite{GKGTJ04}
\begin{equation}
  n(E)dE=p^2dp\int d\Omega_\mathbf{p} \ \left|T_{\p{diss}}(\mathbf{p})\right|^2 \, .
  \label{dissociationspectrumgeneral}
\end{equation}
Here $\int d\Omega_\mathbf{p}$ denotes the integration
over angles, $E=p^2/m$, and $T_{\p{diss}}(\mathbf{p})$ is the transition amplitude, given by
\begin{equation}
T_{\p{diss}}(\mathbf{p}) = \langle\phi_\mathbf{p}^\p{f}|
  U_\mathrm{2B}(t_{\p{f}},t_{\p{i}})|\phi_\mathrm{b}^\p{i}\rangle .  \label{transitionamplitude}
\end{equation}
Here, $\phi_\mathbf{p}^\p{f} = \phi_\mathbf{p}(B_\p{f})$ is the scattering state of momentum $\mathbf{p}$ at the final magnetic field, and $\phi_\p{b}^\p{i} = \phi_\p{b}(B_\p{i})$ is the Fesh\-bach molecular state at the initial magnetic field. In the following, we consider only resonances for which a spherically symmetric Fesh\-bach resonance state is coupled to the entrance channel by spin exchange~\cite{review}. Consequently, the transition amplitude depends only on the modulus of the momentum. 


\section{Dissociation due to a magnetic field jump}
\label{sec:jump}

\begin{figure}[!htbp]
  \begin{center}
    \includegraphics[width=\columnwidth,clip]{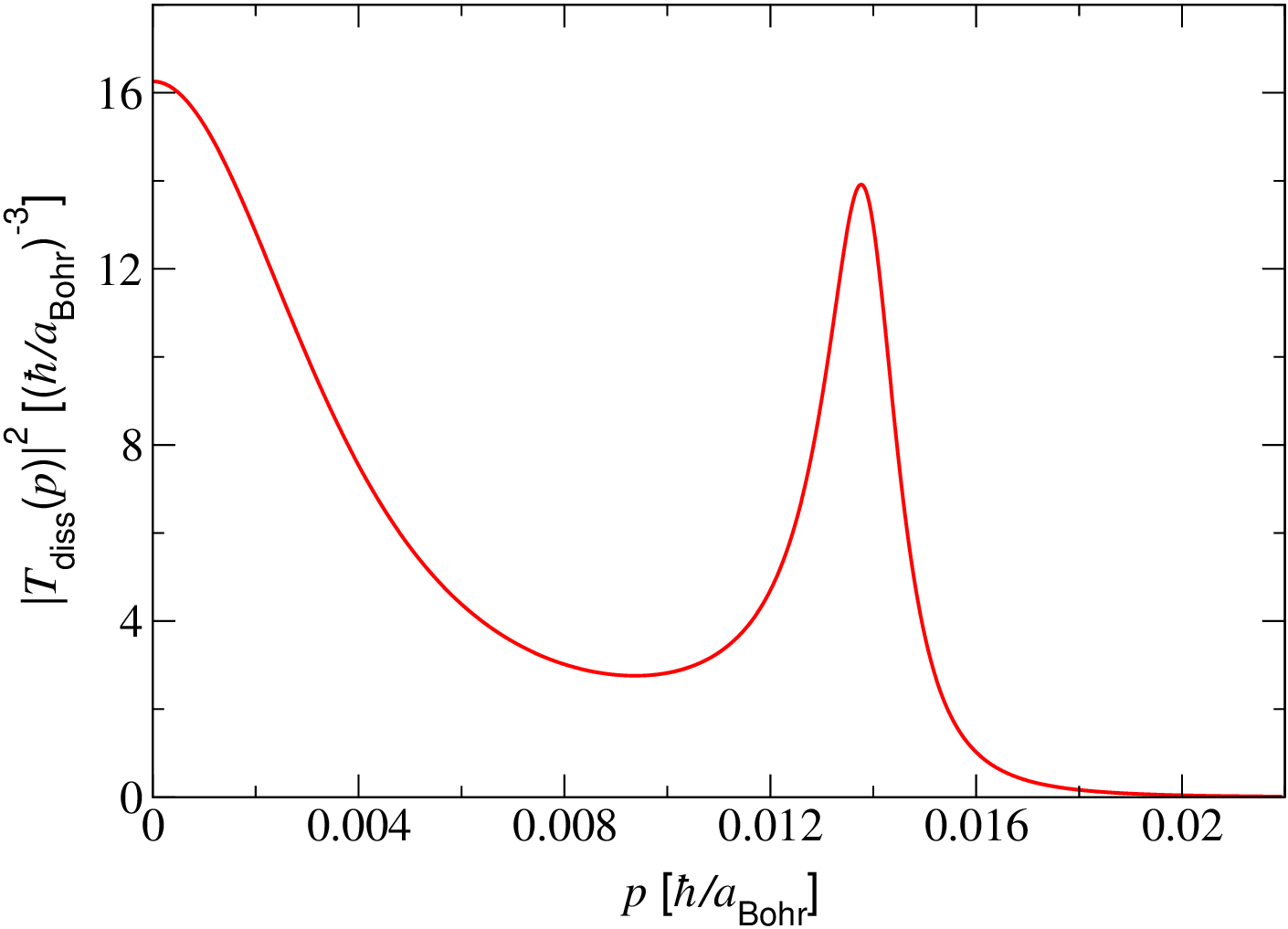}
    \caption{(Color online) Transition probability density $|T_\mathrm{diss}(p)|^2$ for the dissociation of a $^{87}$Rb$_2$ Fesh\-bach molecule into a pair of atoms with relative momentum $p$, due to a jump across the 1007.4\,G resonance. The initial and final fields used here are $B_{\p{i}}=1007$\,G and $B_{\p{f}}=1009.4$\,G. The momentum $p_\p{res}=0.0140\, \hbar/a_\p{Bohr}$, corresponding to the resonance energy $E_\p{res}(B_{\p{f}})=p_\p{res}^{2}/m$ at the end of the jump, gives the approximate position of the peak.}
    \label{fig:2peaks}
  \end{center}
  \end{figure}

In this section we discuss the limit of an infinitely fast ramp across a zero-energy resonance, corresponding to an instantaneous change in the magnetic-field strength. This scenario is directly related to the recent experiment of D{\"u}rr \textit{et al.}~\cite{DuerrPRA04}, where a sharply peaked dissociation spectrum was observed following a fast ramp across the 685\,G resonance of $^{87}$Rb. For a jump, the transition amplitude of Eq.~(\ref{transitionamplitude}) is completely determined by the overlap of the initial Fesh\-bach molecular state and the final scattering state,
\begin{align}
T_\p{diss}(p)\underset{t_{\p{f}} \to t_{\p{i}}}{\longrightarrow}
\langle\phi_\mathbf{p}^\p{f}|\phi_\p{b}^\p{i}\rangle\, .
\end{align}
Evaluating this overlap using the approach of Section~\ref{sec:2channelapproach} reveals that the transition probability density $|T_\p{diss}(p)|^2$ can have a two-peaked structure. An example of such a structure, for the 1007.4\,G resonance of $^{87}$Rb, is shown in Fig.~\ref{fig:2peaks}. This particular form of $|T_\mathrm{diss}(p)|^2$ can be understood by considering the entrance-channel and closed-channel components of $\phi_\mathrm{b}^\p{i}$ and $\phi_\mathbf{p}^\p{f}$, given by Eqs.~(\ref{phib}) and (\ref{phip}) respectively. The maximum at zero momentum is created by the overlap of the entrance-channel components of the initial molecular and final scattering states. The appearance of the peak at nonzero energies is controlled by the amplitude $A(B_{\p{f}},E)$, which determines the final scattering states $\phi_\mathbf{p}^\p{f}$ through Eq.~(\ref{amplitude}). Provided that the influence of $\Delta(E)$ and $\gamma(E)$ in Eq.~(\ref{amplitude}) can be neglected, $|T_\p{diss}(p)|^2$ tends to peak around $E \approx E_\p{res}(B_{\p{f}})$. This is the case for the example of Fig.~\ref{fig:2peaks}.

In general, though, the influence of $\Delta(E)$ and $\gamma(E)$ can be significant, leading to the broadening and possibly even disappearance of the nonzero-energy peak in the dissociation spectrum. For small $E = (\hbar k)^{2}/m$, satisfying $|ka| \ll 1$, the two-body interactions can be completely characterised in terms of the scattering length. The regime where the physics depends only on the scattering length is referred to as the Wigner threshold law regime for positive energies, and as the universal regime for negative energies. The energy shift $\Delta(E)$ assumes its maximum value for $E \to 0$,
\begin{align}
\Delta(E)  \underset{k \to 0}{\sim}
\mu_\mathrm{res}(B_\mathrm{res}-B_0)\, .
\label{Delta}
\end{align}
In this limit $\Delta(E)$ is related to the quantity $B_0 - B_{\p{res}}$, which is termed the resonance shift. The decay width $\gamma(E)$ in the Wigner threshold law regime assumes the classic form of~\cite{WignerPR48}
\begin{align}
\gamma(E) \underset{k\to 0}{\sim} k a_\mathrm{bg} \mu_\mathrm{res}
\Delta B \, ,
\label{gamma}
\end{align}
where $\Delta B$ is the resonance width, and $a_{\p{bg}}$ is the background scattering length. The resonance width and background scattering length are related to the scattering length $a$ by 
\begin{align}
a(B) = a_{\p{bg}}\left(1 - \frac{\Delta B}{B - B_0}\right) \, .
\label{eq:a}
\end{align}
Equation~(\ref{gamma}) indicates that $\gamma(E)$ can be significant for broad resonances with large $|a_{\p{bg}}|$, which will therefore have broad peaks in their dissociation spectra. The modulus of the resonance shift $B_0-B_{\p{res}}$ of Eq.~(\ref{Delta}) also tends to be large for these resonances. A derivation beyond the scope of this paper shows that the resonance shift is given by \cite{GKGTJ04, jul89}
\begin{align}
B_0-B_\mathrm{res}=\Delta B\frac{a_\mathrm{bg}}{\bar{a}}
   \left[
     \frac{1-a_\mathrm{bg}/\bar{a}}{1+\left(1-a_\mathrm{bg}/\bar{a}\right)^2}
   \right].
   \label{magicformula}
\end{align}
Here, $\bar{a}$ is the mean scattering length given by~\cite{gri93}
\begin{align}
\bar{a}=\frac{\Gamma(3/4)}{\sqrt{2}\Gamma(5/4)}l_\mathrm{vdW} \, ,
\end{align}
and $l_\mathrm{vdW}=\frac{1}{2}(mC_6/\hbar^2)^{1/4}$ is the van der Waals length, characterising the long-range part of the entrance-channel interaction potential according to $V_{\p{bg}}(r)\underset{r\to\infty}{\sim}-C_6/r^6$. 

The defining property of entrance-channel dominated resonances is that the universal formula for the bound state energy, $E_\mathrm{b}\approx -\hbar^2/ma^{2}$, is accurately extended beyond the universal regime by \mbox{$E_\mathrm{b}\approx -\hbar^2/[m(a - \bar{a})^2]$}~\cite{review, gri93}. The accuracy of this extension depends on the degree to which the influence of the van der Waals tail of the entrance-channel potential dominates over that of the Fesh\-bach resonance state~\cite{KGG04}. Conversely, if the influence of the Fesh\-bach resonance state is dominant, the resonance is referred to as closed-channel dominated. The distinction between entrance-channel and closed-channel dominated resonances is quantified by~\cite{stoll05, petrov04}
\begin{align}
  \eta=
  \frac{\bar{a}}{a_\mathrm{bg}}
  \frac{\hbar^2/(m\bar{a}^2)}{\mu_\mathrm{res}\, \Delta B},
  \label{eta}
\end{align}
which we term the closed-channel dominance. For entrance-channel dominated resonances the relation $\eta\ll 1$ is fulfilled, implying that $|a_\mathrm{bg}|$ and/or $|\mu_\mathrm{res} \Delta B|$ are large. The latter quantity being large indicates strong interchannel coupling~\cite{GKGTJ04}. This leads to the closed-channel admixture of the Fesh\-bach molecular state being suppressed over a range of magnetic fields that is wider than the universal regime. Consequently, the wavefunction has a long-range halo structure~\cite{review} within this range of fields. 

Although the closed-channel dominance of Eq.~(\ref{eta}) was derived in the context of the properties of the Fesh\-bach molecular state, it can be probed experimentally by measuring dissociation spectra. This can be seen by analysing the position of the finite-energy peak, $E_\mathrm{peak}$, relative to its spectral width, $\Delta E_\mathrm{peak}$. By using the Wigner threshold law expressions of Eqs.~(\ref{Delta}) and (\ref{gamma}), it is possible to show that the finite-energy peak is approximately Lorentzian. Consequently, $\Delta E_\mathrm{peak}$ will be taken as its half-width at half-maximum. A distinct peak appears in the jump dissociation spectrum when $E_\mathrm{peak}/\Delta E_\mathrm{peak} \gg 1$. Evaluating this quantity in the Wigner threshold law regime yields the expression for the peak clarity,
\begin{align}
\frac{E_\mathrm{peak}}{\Delta E_\mathrm{peak}}=\eta \frac{
  \mu_\mathrm{res}(B_{\p{f}}-B_0)/E_\mathrm{vdW}-1/(2\eta^2)}
     {\sqrt{\mu_\mathrm{res}(B_{\p{f}}-B_0)/E_\mathrm{vdW}-1/(4\eta^2)}}\, ,
     \label{peakness}
\end{align}
where we refer to $E_\mathrm{vdW}=\hbar^2/m\bar{a}^2$ as the van der Waals energy. $E_\mathrm{vdW}$, which greatly exceeds the relative kinetic energies typical of cold atom pairs, gives the order of magnitude of the spacing between the highest excited vibrational bound states. It therefore sets the largest energy scale able to be experimentally accessed without other resonance phenomena arising, and limits the range of energies able to be considered within a single-resonance approach. In the range of validity $\mu_{\p{res}}(B_{\p{f}} - B_{0}) \leq E_{\p{vdW}}$, Eq.~(\ref{peakness}) gives the inequality
\begin{align}
\frac{E_\mathrm{peak}}{\Delta E_\mathrm{peak}} \leq \eta \frac{1-1/(2\eta^2)}
     {\sqrt{1-1/(4\eta^2)}}\, ,
     \label{peaknesslim}
\end{align}
with equality for $\mu_{\p{res}}(B_{\p{f}} - B_{0}) = E_{\p{vdW}}$. It is evident from Eq.~(\ref{peaknesslim}) that the condition for the existence of a distinct peak at a nonzero energy is $\eta \gg 1$, which is characteristic of closed-channel dominated resonances. An example of a resonance falling into this category is provided by the 1007.4\,G resonance of $^{87}$Rb, for which $\eta=5.9$ (see Fig.~\ref{fig:2peaks}). Entrance-channel dominated resonances, characterised by $\eta \ll 1$, exclude the possibility of a clear peak appearing at a nonzero energy in the dissociation spectrum. The presence or absence of a sharp peak in the dissociation spectrum is therefore an indicator of whether a zero-energy resonance is entrance-channel or closed-channel dominated.

\begin{figure}[!htbp]
  \begin{center}
    \includegraphics[width=\columnwidth,clip]{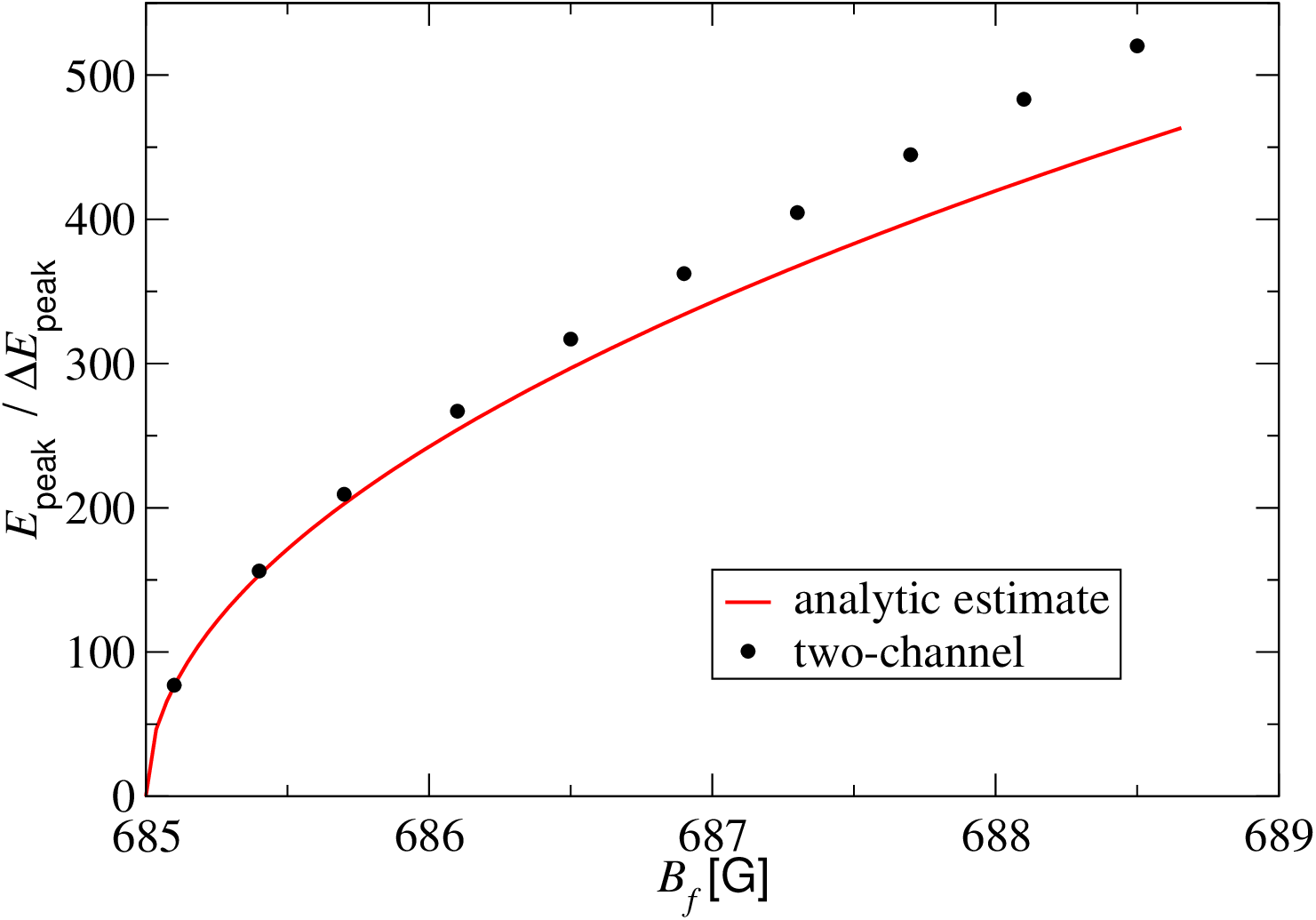}
    \caption{(Color online) The measure of peak clarity $E_\mathrm{peak}/\Delta E_\mathrm{peak}$ in the dissociation spectrum of a $^{87}$Rb$_2$ molecule, using the 685\,G resonance. This resonance has a high closed-channel dominance of $\eta = 463$. The initial field used is $B_{\p{i}}=684.5$\,G. The dots indicate values extracted from fits to the spectra obtained from full two-channel single-resonance calculations, while the solid line shows the approximate analytic predictions of Eq.~(\ref{peakness}). At $B\sim 688$\,G, the single-resonance approach breaks down as the energy range covered starts to exceed the van der Waals energy  $E_\mathrm{vdW}$.}
    \label{fig:comparison}
  \end{center}
  \end{figure}

We have verified the quality of the estimate given by Eq.~(\ref{peakness}) by comparing it to the results of calculations using the approach of Section~\ref{sec:2channelapproach}. The comparison, shown in Fig.~\ref{fig:comparison}, confirms that Eq.~(\ref{peakness}) provides a good estimate of the ratio $E_\mathrm{peak}/\Delta E_\mathrm{peak}$. Moreover, the analytic expression of  Eq.~(\ref{peakness}) slightly underestimates $E_\mathrm{peak}/\Delta E_\mathrm{peak}$, meaning that the peak clarity one would expect to see experimentally is always slightly greater than the analytic estimate.

\begin{figure}[!htbp]
  \begin{center}
    \includegraphics[width=\columnwidth,clip]{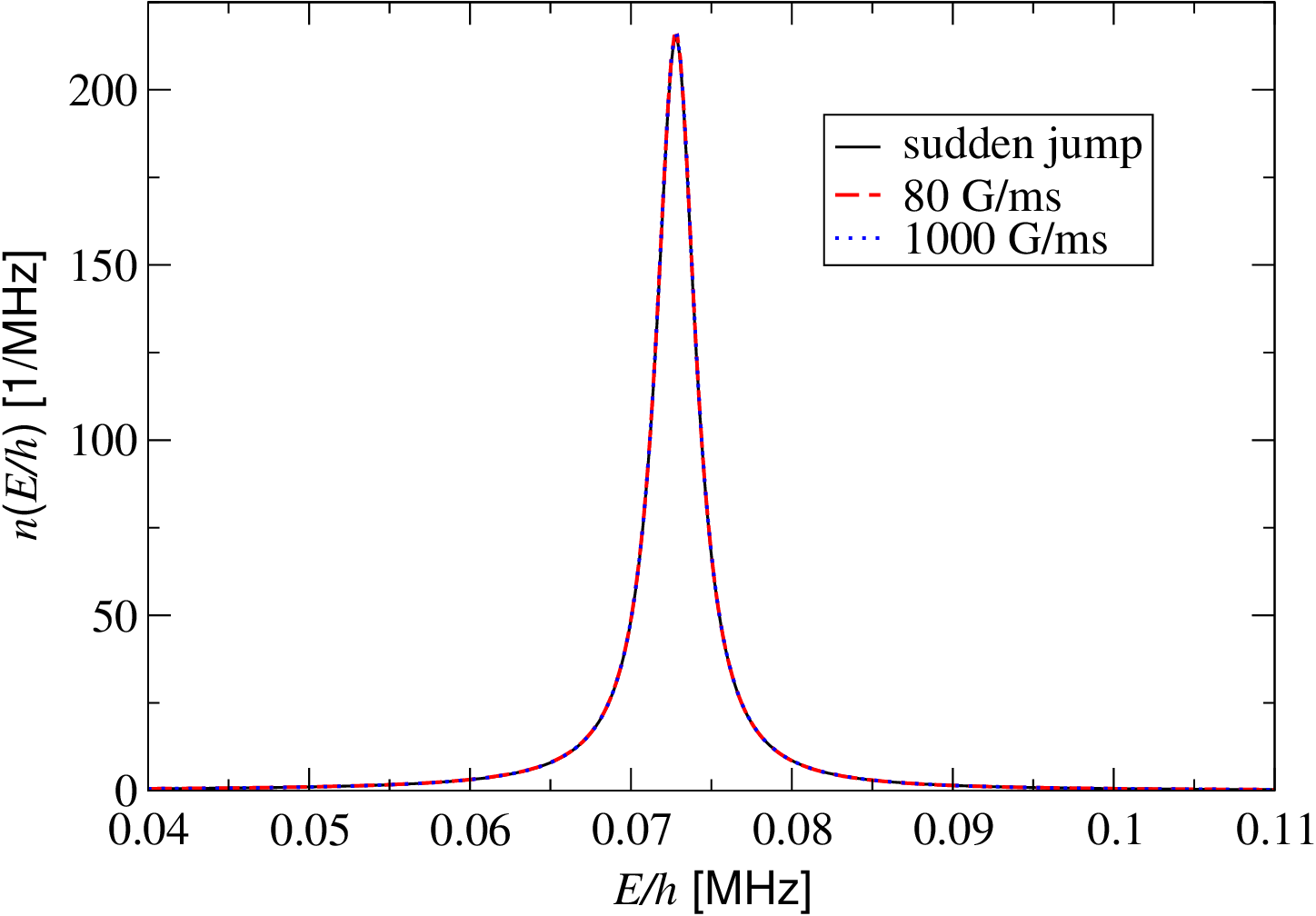}
    \caption{(Color online) Dissociation spectrum of $^{87}$Rb$_2$ molecules produced in the vicinity of the $B_0=685$\,G zero-energy resonance. The initial and final magnetic field strengths are $B_{\p{i}}=B_0\,-\,$50\,mG \cite{Duerrprivate} and $B_{\p{f}}=B_0\,+\,$40\,mG \cite{DuerrPRA04}. 80\,G/ms is the ramp speed reported in Ref.~\cite{DuerrPRA04}, while 1000\,G/ms is approximately the maximum currently available~\cite{Duerrprivate}. The distinct peak is centred around $h\,\times\,$0.073 MHz, corresponding to 3.5\,$\mu$K, as compared to the 3.3\,$\mu$K that can be deduced from Ref.~\cite{DuerrPRA04}.
 }
    \label{fig:Munich}
  \end{center}
  \end{figure}

The preceding analysis treats the ideal case of an instantaneous change of the magnetic field strength. An experiment is described in Ref.~\cite{DuerrPRA04} that actually operates in this regime. After an 80\,G/ms ramp of the magnetic field across the 685\,G resonance of $^{87}$Rb, the atomic gas was found to expand outwards as a spherical shell, corresponding to a sharply peaked dissociation spectrum. In order to assess how closely the jump approximation follows the experimental observations of Ref.~\cite{DuerrPRA04}, we have performed an exact two-channel calculation of the dissociation spectrum of Eq.~(\ref{dissociationspectrumgeneral}), for the experimental ramp speed of 80\,G/ms. The comparison is shown in Fig.~\ref{fig:Munich}, and confirms that the 80\,G/ms ramp is fast enough to produce a spectrum virtually indistinguishable from that of a jump. Consequently, the experiment of Ref.~\cite{DuerrPRA04} provides a clear probe of the closed-channel admixtures of the initial bound and final scattering states, and of the strong closed-channel dominance ($\eta = 463$) of this particular resonance. We note that experiments reported in Ref.~\cite{DuerrPRA04} using the less closed-channel dominated ($\eta = 5.9$) 1007\,G resonance of $^{87}$Rb did not produce such a peak for the 80\,G/ms ramp-speed used. We have calculated that to be in the jump regime for this resonance would require a ramp speed of 5000\,G/ms or above. The minimum ramp speed to be in the regime of a jump is different for each resonance, however ramp speeds of order 1000\,G/ms are now possible~\cite{Duerrprivate}.

\begin{figure}[!htbp]
  \begin{center}
    \includegraphics[width=\columnwidth,clip]{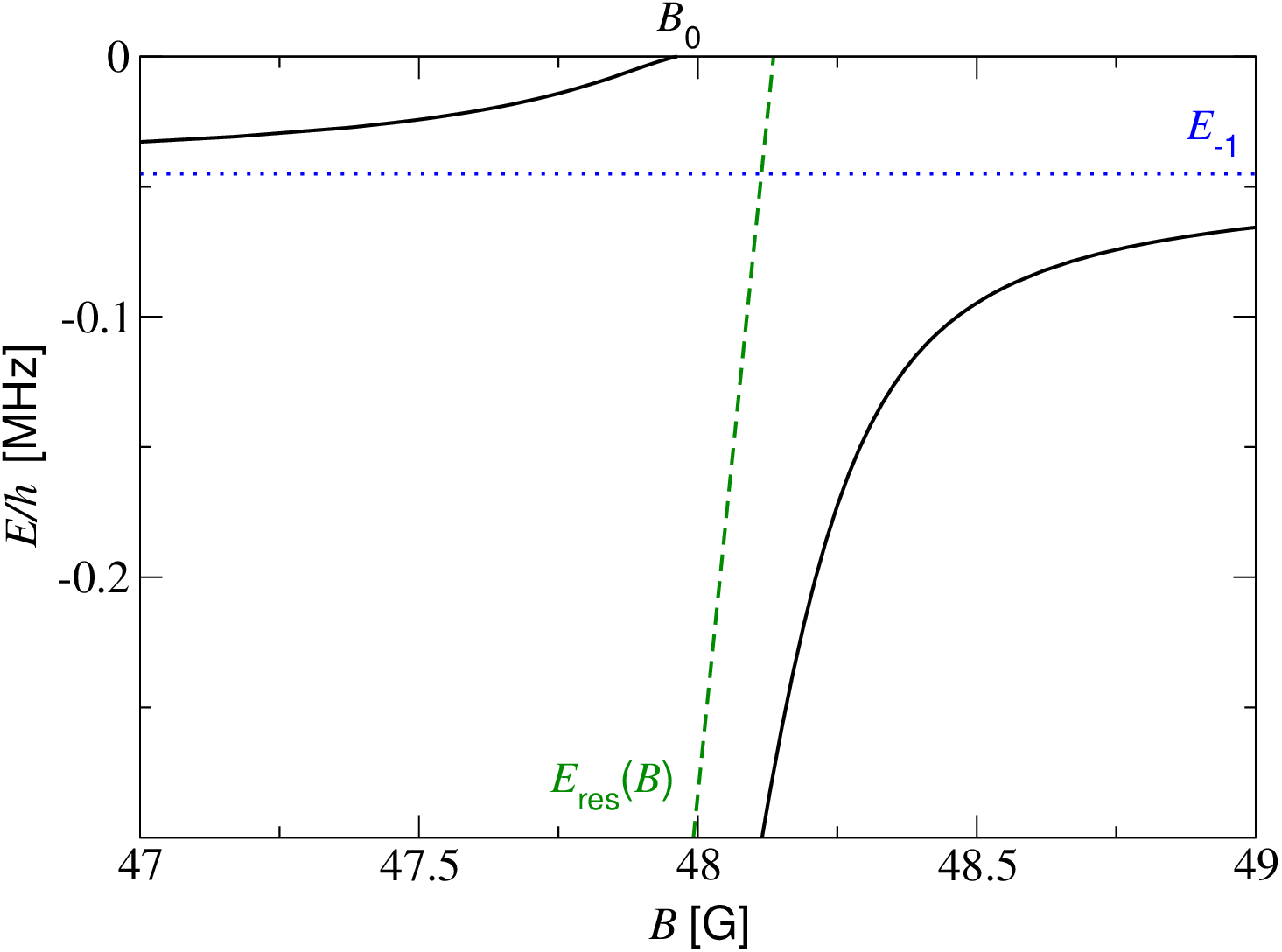}
    \caption{(Color online) Energies of the two highest excited vibrational bound states near the 48\,G resonance of $^{133}$Cs. For fields far below the resonance, the Fesh\-bach molecular state adiabatically correlates with the highest excited bound state $\phi_{-1}$ of the entrance channel (dotted line), while the next deeper bound state corresponds to the Fesh\-bach resonance state (dashed line). Above $B_0$ the Fesh\-bach molecule does not exist and the next deeper state converges to $\phi_{-1}$. Due to this avoided crossing being narrow, the possible transfer into the final bound state from a jump can be as large as 10\%.}
    \label{fig:avoided}
  \end{center}
  \end{figure}

At magnetic fields far from $B_0$, the highest excited vibrational bound state of $H_{\p{2B}}(B)$ often corresponds to the highest excited vibrational state of the entrance channel potential, $\phi_{-1}$, as illustrated in Fig.~\ref{fig:avoided}. When the Fesh\-bach molecular state is subjected to a magnetic field sweep across the resonance, some population can then be transferred to the highest excited vibrational bound state at the final field strength. Some Fesh\-bach molecules will therefore not be dissociated during a ramp. We have calculated this effect for the 685\,G resonance of $^{87}$Rb and found it to be negligible. For the 48\,G resonance of $^{133}$Cs referred to in Fig.~\ref{fig:avoided}, however, we have found that the transfer into the bound state can be as much as 10\,\%. In our calculations we have assumed sweep rates of no more than 1000\,G/ms, which is currently achievable experimentally~\cite{Duerrprivate}. The transfer is unusually large for this particular resonance because the entrance-channel potential supports a state with an energy of only $E_{-1} = -h \times 45\,$kHz~\cite{review}. This makes the avoided crossing between the bound states narrow, as shown in Fig.~\ref{fig:avoided}, and so the wave functions of the Fesh\-bach molecule and the final bound state converge to $\phi_{-1}$ quickly as $B$ moves away from $B_0$. This increases the overlap between the initial and final bound states and so makes the transfer from a jump larger.

  
\section{Transition from jumps to asymptotic ramps}
\label{sec:asymptotic}

\begin{figure}[!htbp]
  \begin{center}
    \includegraphics[width=\columnwidth,clip]{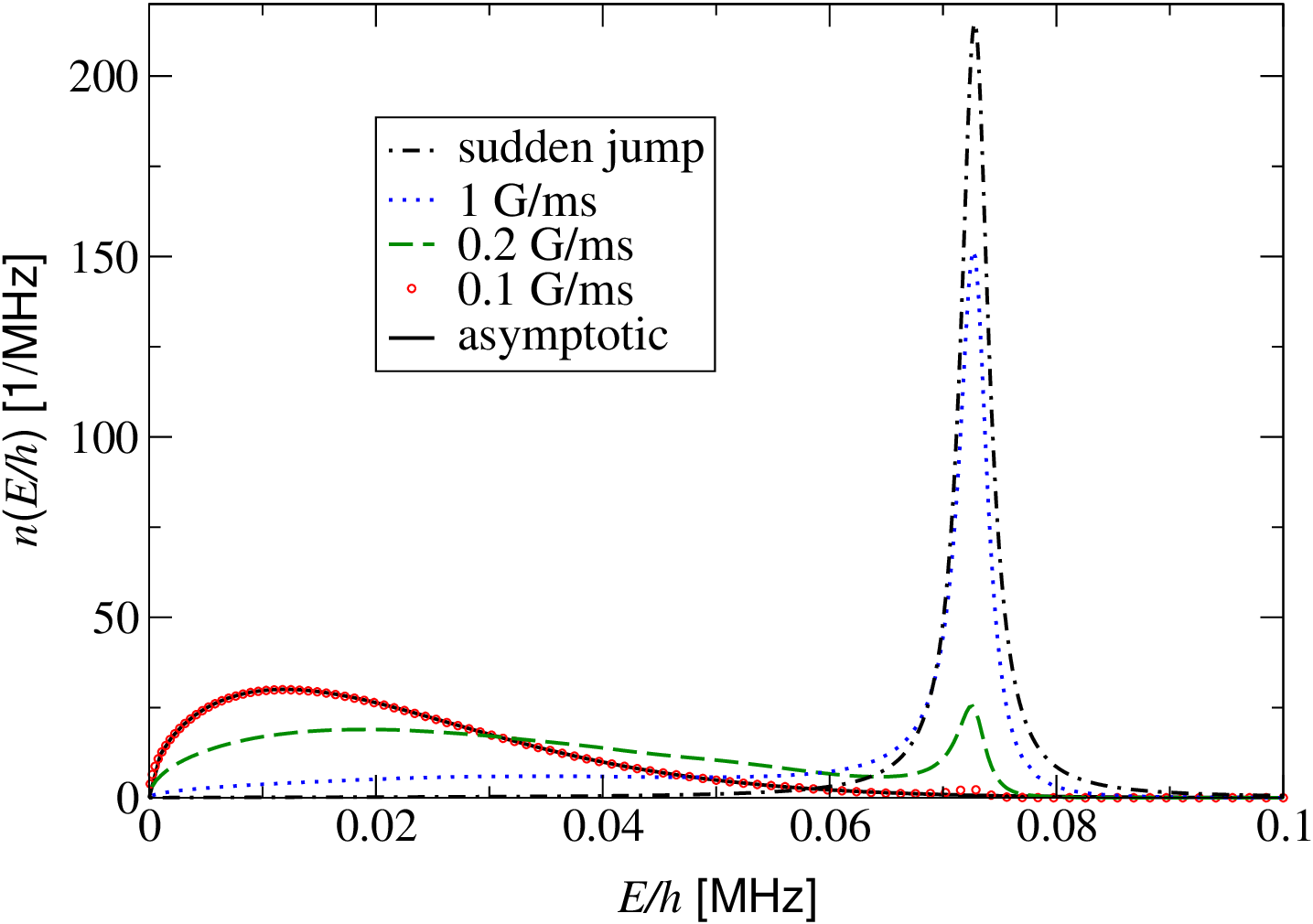}
    \caption{(Color online) A sequence of dissociation spectra of $^{87}$Rb$_2$       molecules, produced near the $685$\,G zero-energy resonance, for a range of ramp speeds. The initial and final magnetic field strengths are $B_{\p{i}}=B_0\,-\,50$\,mG and $B_{\p{f}}=B_0\,+\,40$\,mG, respectively, and correspond to the conditions of Ref.~\cite{DuerrPRA04}. The dash-dotted curve corresponds to a jump of the magnetic field strength and is identical to the curves shown in Fig.~\ref{fig:Munich}. The circles illustrate the spectrum obtained from an exact two-channel calculation for $\dot{B}=0.1$\,G/ms, and coincides with the evaluation of Eq.~(\ref{finalspectraldensity}) indicated by the superimposed solid line.
 }
    \label{fig:progression}
  \end{center}
  \end{figure}

In this section we consider ramp speeds that are intermediate between the jumps of Section~\ref{sec:jump} and asymptotic ramps. Keeping the initial and final magnetic fields shown in Fig.~\ref{fig:Munich} for all ramp speeds,  we calculate the dissociation spectra including the full time evolution of Eq.~(\ref{transitionamplitude}). The resulting spectra have the characteristic double-peaked structure for jumps, but coincide with the asymptotic form for sufficiently slow ramps. This transition is shown in Fig.~\ref{fig:progression}. For any given ramp speed, however, making the ramp wide enough will result in a spectrum of the asymptotic form. 

For linear ramps of the form of Eq.~(\ref{eq:bramp}), it is possible to derive an exact formula for the time-evolution operator $U_{\p{2B}}(t,t')$ associated with the two-body Hamiltonian $H_{\p{2B}}(B(t))$. Here we consider a continuum of positive energies as represented by the scattering states of Eq.~(\ref{phip}), in contrast to the set of discrete levels examined in previous analytic treatments~\cite{review, dem68, mac98}. We separate Eq.~(\ref{H2BofB}) into its stationary and explicitly time dependent parts,
\begin{align}
  H_\mathrm{2B}(B(t))=H_\mathrm{stat}+H_\mathrm{cl}(t) \, ,
  \label{H2Boft}
\end{align}
where $H_\mathrm{stat}=H_\mathrm{2B}(B_\mathrm{res})$, and we now express the single-resonance, closed-channel Hamiltonian of Eq.~(\ref{eq:hcl}) as a function of time, i.e. $H_{\p{cl}}(B(t)) \to H_{\p{cl}}(t)$. Using the separability of $H_{\p{cl}}(t)$, the derivation of Appendix~\ref{app:EvolutionOperator} leading to Eq.~(\ref{G2B}) shows that $U_\p{2B}(t,t')$ can be calculated from only its closed-channel component,
\begin{align}
g_\mathrm{2B}^{(+)}(t,t')=\frac{1}{i\hbar}\theta(t-t')\langle\phi_\p{res},\mathrm{cl}|
  U_\p{2B}(t,t')|\phi_\p{res},\mathrm{cl}\rangle \, .
  \label{g2b}
\end{align}
Here, $\theta(t-t')$ is the step function, which yields unity for \mbox{$t - t' > 0$} and 0 elsewhere. The linear variation of the resonance energy makes it possible to represent $g_\mathrm{2B}^{(+)}(t,t')$ in terms of the following Fourier integrals:
\begin{align}
  g_\mathrm{2B}^{(+)}(t,t')=\int dE
  \frac{e^{-iE(t-t')/\hbar}}{2\pi\hbar}\int_E^\infty dE'
  \frac{e^{i[\varphi(z',t')-\varphi(z,t')]}}{i\hbar\dot{E}_\mathrm{res}}\, .
  \label{g2bfourier}
\end{align}
Here $z = E + i0$, $z' = E' + i0$, and the phase $\varphi(z,t)$ is given in Eq.~(\ref{phase}). This leads to the expression of Eq.~(\ref{G2Boftexact}) for the time-evolution operator. Consequently, $U_\mathrm{2B}(t_{\p{f}},t_{\p{i}})$ is expressed solely in terms of the known energy-dependent Green's function $G_\mathrm{stat}(z)$ associated with the stationary Hamiltonian $H_\mathrm{stat}$ and its closed channel matrix element $g_\mathrm{stat}(z)$. 

In Appendix~\ref{app:EvolutionOperator} we use Eq.~(\ref{G2Boftexact}) together with a stationary phase analysis to evaluate the transition amplitude of Eq.~(\ref{transitionamplitude}) in the limit of an asymptotically wide ramp. This shows the transition probability density to be
\begin{align}
|T_\mathrm{diss}(p)|^2=&\frac{2\pi}{\hbar|\dot{E}_\mathrm{res}|} |\langle\phi_\mathbf{p}^{(+)}|W|\phi_\mathrm{res}\rangle|^2 \, \times \nonumber \\
&\times\exp\left(-\frac{2\pi}{\hbar|\dot{E}_\mathrm{res}|}
  \int d\mathbf{p'}
 \theta(p-p')|\langle\phi_\mathbf{p'}^{(+)}|W|\phi_\mathrm{res}\rangle|^2\right)
  \, .\label{probabilityexact}
\end{align}
Under typical experimental conditions the energies of the dissociated atoms do not exceed several $\mu$K, which is usually well within the Wigner threshold law domain. Consequently, the interchannel coupling matrix elements in Eq.~(\ref{probabilityexact}) can be evaluated in the limit of zero relative momentum and related to the resonance width $\Delta B$ through \cite{GKGTJ04}
\begin{align}
  \Delta B=\frac{m}{4\pi\hbar^2 a_\mathrm{bg}}
  \frac{(2\pi\hbar)^3
    \left|\langle\phi_\mathrm{res}|W|\phi_0^{(+)}\rangle\right|^2}
       {\mu_\mathrm{res}} \, .
       \label{resonancewidth}
\end{align}
This constitutes a rigorous derivation of the asymptotic dissociation spectrum,
\begin{equation}
  n(E)=-\frac{\partial}{\partial E}
  \exp\left(-\frac{4}{3}\sqrt{\frac{mE}{\hbar^2}}\left|a_\mathrm{bg}\right|
  \frac{E|\Delta B|}{\hbar\left|\dot{B}\right|}\right),
  \label{finalspectraldensity}
\end{equation}
which was previously inferred in Refs.~\cite{MukaiyamaPRL04, GKGTJ04} from a Fermi Golden Rule argument. The accuracy of this expression was confirmed by the experiments of Mukaiyama {\it et al.}~\cite{MukaiyamaPRL04} with $^{23}$Na and D{\"u}rr {\it et al.}~\cite{DuerrPRA04} with $^{87}$Rb. This led to the application of Eq.~(\ref{finalspectraldensity}) to measuring the widths of previously unexplored resonances \cite{DuerrPRA04}. The dissociation spectrum of Eq.~(\ref{finalspectraldensity}) is plotted in Fig.~\ref{fig:progression} as the asymptotic limit of slow ramps for the 685\,G resonance of $^{87}$Rb. 

The dissociation spectrum in Eq.~(\ref{finalspectraldensity}) does not depend on the magnetic moment $\mu_\mathrm{res}$ of the resonance state, but just on $|a_{\p{bg}}\Delta B|$. This product characterises the scattering length in the close vicinity of a zero-energy resonance according to $a\approx - a_\mathrm{bg} \Delta B/(B-B_0)$. This suggests that it should be possible to arrive at Eq.~(\ref{finalspectraldensity}) using different descriptions of the underlying two-body physics, the only condition being that the near-resonant scattering properties are correctly recovered. Single-channel descriptions~\cite{KGB03,KGG04,SzymanskaPRA05} are capable of describing entrance-channel dominated resonances, such as the 48\,G resonance of $^{133}$Cs shown in Fig.~\ref{fig:avoided}, over a wide range of magnetic fields. The approach of Appendix~\ref{app:SingleChannel} varies the potential in such a way that the scattering length $a$ is recovered for all magnetic fields, as well as $E_{-1}$, the energy of the highest excited vibrational entrance-channel state away from the resonance. For closed-channel dominated resonances, however, single-channel approaches are valid only in the largely inaccessible universal regime. Taking a separable form of the potential allows Eq.~(\ref{finalspectraldensity}) to be derived analytically, as shown in Appendix~\ref{app:SingleChannel}. Consequently, for asymptotic ramps the physical description need only apply to the near resonant region, whereas for jumps it must apply over the entire range of magnetic fields. 

\section{Conclusions}
\label{sec:conclusion}

We have studied in detail the dissociation spectra produced by linear sweeps of the magnetic field strength across zero-energy resonances, for field ranges and ramp speeds varying from jumps to asymptotic ramps. A key finding of our studies is that jumps in the magnetic field can be used to classify zero-energy resonances according to the dominance of their entrance and closed channels. Our scheme requires measuring whether a sharp peak occurs in the dissociation spectrum near the final resonance energy. The presence or absence of this peak allows a resonance to be classified as closed-channel or entrance-channel dominated, respectively. This may be measured before other properties of the resonance are studied. By comparing the spectrum of a jump to exact numerical calculations for the example of the 685\,G resonance of $^{87}$Rb, we have shown that the ramp speed required for such a measurement is well below the maximum currently possible. 

Our calculations of the dissociation spectrum for varying ramp speeds and fixed initial and final fields show that the spectrum approaches the asymptotic form as the ramp speed is lowered. Increasing the width of the ramp for a fixed ramp speed also brings the spectrum closer to the asymptotic result. We have analytically determined the time evolution for linear ramps using a two-channel approach. Evaluating this in the asymptotic limit rigorously produces the asymptotic spectrum of Eq.~(\ref{finalspectraldensity}). Asymptotic ramps, however, only require the physical description to correctly describe the near-resonant region, and so may also be described using a single-channel approach.

\section{Acknowledgements}
This research has been supported by the General Sir John Monash
Foundation and Universities UK (T.M.H.), the Royal Society (K.G. and T.K.), and the ESF Quantum
Degenerate Dilute Systems (QUDEDIS) Programme (E.W.). We are grateful to K. Burnett, S. D\"urr and J. I. G\'oral for interesting discussions.


\begin{appendix}
\section{Two-channel time evolution operator}
\label{app:EvolutionOperator}

In this appendix we derive an exact expression for the time evolution operator in the two channel, single resonance approach. We then use this result to evaluate the transition amplitude for the case of an asymptotic ramp, which provides a rigorous derivation of the corresponding dissociation spectrum. We calculate the transition amplitude in terms of the retarded Green's function $G_\mathrm{2B}^{(+)}(t,t')$, related to the time evolution operator $U_\mathrm{2B}(t,t')$ by
\begin{align}
G_\mathrm{2B}^{(+)}(t,t')=\frac{1}{i\hbar}\theta(t-t')U_\mathrm{2B}(t,t')
\, .
\label{G2Boftdefinition}
\end{align}  
We separate the constant and time-dependent parts of the Hamiltonian as in Eq.~(\ref{H2Boft}). The retarded
Green's function associated with the stationary Hamiltonian
$H_\mathrm{stat}$ in Eq.~(\ref{H2Boft}) is given by
\begin{align}
G_\p{stat}^{(+)}(t-t') &= \frac{1}{i\hbar}\theta(t-t')
\exp[-iH_\mathrm{stat}(t-t')/\hbar] \, .
\end{align}
The separation of $H_\p{2B}(B)$ into stationary and time-dependent parts then gives two equivalent integral equations for the full retarded Green's function:
\begin{align}
  G_\mathrm{2B}^{(+)}(t,t')&=G_\mathrm{stat}^{(+)}(t-t')
  +\int d\tau \,G_\mathrm{2B}^{(+)}(t,\tau)H_\mathrm{cl}(\tau)
  G_\mathrm{stat}^{(+)}(\tau-t') \, ,
\label{postG2B}
\end{align}
referred to as the preform, and
\begin{align}
  G_\mathrm{2B}^{(+)}(t,t')&=G_\mathrm{stat}^{(+)}(t-t')
  +\int d\tau \,G_\mathrm{stat}^{(+)}(t-\tau)H_\mathrm{cl}(\tau)
  G_\mathrm{2B}^{(+)}(\tau,t') \, , 
  \label{preG2B}
\end{align}
termed the postform. Differentiating either form of the integral equation with respect to $t$ and using Eq.~(\ref{G2Boftdefinition}) recovers Eq.~(\ref{eq:U}) defining $U_\mathrm{2B}(t,t')$. Substituting Eq.~(\ref{postG2B}) into Eq.~(\ref{preG2B}) and using the form of $H_\mathrm{cl}(t)$ in Eq.~(\ref{eq:hcl}) gives the following general expression for the two-body time-dependent Green's function:
\begin{align}  G&_\mathrm{2B}^{(+)}(t_{\p{f}},t_{\p{i}})=G_\mathrm{stat}^{(+)}(t_{\p{f}}-t_{\p{i}}) \, + \nonumber \\
  +&\int d\tau \,G_\mathrm{stat}^{(+)}(t_{\p{f}}-\tau)|\phi_\mathrm{res},\mathrm{cl}\rangle
E_\mathrm{res}(\tau) \bigg[
\langle\phi_\mathrm{res},\mathrm{cl}|G_\mathrm{stat}^{(+)}(\tau-t_{\p{i}}) \, + \nonumber \\
+& \int d\tau' \, g_\mathrm{2B}^{(+)}(\tau,\tau')
  E_\mathrm{res}(\tau')   \langle\phi_\mathrm{res},\mathrm{cl}|G_\mathrm{stat}^{(+)}(\tau'-t_{\p{i}})\bigg]\, .
\label{G2B}
\end{align}
Here, $g_\mathrm{2B}^{(+)}(t,t')$ is given in Eq.~(\ref{g2b}). Throughout these appendices we use the notation of a lower-case $g$ to refer to the closed-channel matrix element of the corresponding Green's function. 

The explicit form of  $G_\mathrm{2B}^{(+)}(t_{\p{f}},t_{\p{i}})$ in Eq.~(\ref{G2B}) shows that obtaining its closed-channel matrix element 
$g_\mathrm{2B}^{(+)}(t,t')$ is sufficient for fully determining the time evolution. To this end, we project Eq.~(\ref{preG2B}) onto $|\phi_\mathrm{res},\mathrm{cl}\rangle$ from the left and right and again use the definition of $H_{\p{cl}}(t)$ to yield
\begin{align}
  g_\mathrm{2B}^{(+)}(t,t')=g_\mathrm{stat}^{(+)}(t-t')
  +\int d\tau \,g_\mathrm{stat}^{(+)}(t-\tau)E_\mathrm{res}(\tau)
  g_\mathrm{2B}^{(+)}(\tau,t') \, .
  \label{g2Boft}
\end{align}
Performing a Fourier transform of both sides of Eq.~(\ref{g2Boft})
with respect to $t$ leads to the time dependence of $E_{\p{res}}(t)$, given in Eq.~(\ref{Eresoft}), being converted into a derivative with respect to energy, yielding the following ordinary differential equation:
\begin{align}
  i\hbar\frac{\partial}{\partial E}g_\mathrm{2B}(z,t')=
  \hbar\left[\frac{\partial}{\partial E}\varphi(z,t')\right] g_\mathrm{2B}(z,t')+
  1/\dot{E}_\mathrm{res}\, .
  \label{ode}
\end{align}
Here the energy and time dependent matrix element $g_\mathrm{2B}(z,t')$ is given by a Fourier transform with respect to time, $g_\mathrm{2B}(z,t')=\int dt\,e^{iz(t-t')/\hbar}  g_\mathrm{2B}^{(+)}(t,t')\, .$ The argument \mbox{$z=E+i0$} has the dimension of energy and its imaginary part provides the appropriate convergence factor. The energy and time dependent phase $\varphi(z,t')$ is given by
\begin{align}
  \varphi(z,t')=
  \frac{E(t'-t_\mathrm{res})}{\hbar} - \frac{1}{\hbar\dot{E}_\mathrm{res}}\int_0^E\frac{dE'}{g_\mathrm{stat}(z')}\, ,
  \label{phase}
\end{align}
where $z'=E'+i0$, and $g_\mathrm{stat}(z)=\int dt\,e^{iz(t-t')/\hbar}g_\mathrm{stat}^{(+)}(t-t')$. A solution of Equation~(\ref{ode}) can be inferred from the derivative with respect to energy of $e^{i\varphi(z,t')}g_{\p{2B}}(z,t')$. Calculating the inverse Fourier transform of the relation for $g_{\p{2B}}(z,t')$ thus obtained gives Eq.~(\ref{g2bfourier}).

Substituting $g_{\p{2B}}(t,t')$ into Eq.~(\ref{G2B}) leaves the resonance energy and the phase terms as the only explicitly time dependent quantities. The time-dependence due to the resonance energy may be eliminated by writing 
\begin{align}
&(t - t_\p{res})e^{i\varphi(z,t)} = -i\hbar e^{i\varphi(z,t_{\p{res}})}\frac{\partial}{\partial E} e^{iE(t-t_{\p{res}})/\hbar}\, ,
\end{align}
and calculating the partial integrals with respect to energy. The time integrals in Eq.~(\ref{G2B}) may then be expressed as Fourier transforms, reducing the general expression for the two-body retarded Green's function to
\begin{align}
G_\mathrm{2B}^{(+)}&(t_{\p{f}},t_{\p{i}})=\int
\frac{dE}{2\pi\hbar}G_\mathrm{stat}(z)\, \times \nonumber \\
&\times\Bigg\{ e^{-iE(t_{\p{f}}-t_{\p{i}})/\hbar} 
\biggl[1- |\phi_\mathrm{res},\mathrm{cl}\rangle \frac{1}{g_\mathrm{stat}(z)}
\langle\phi_\mathrm{res},\mathrm{cl}|G_\mathrm{stat}(z)\biggr] \,+ \nonumber \\
  &+\int_{-\infty}^{E} dE' 
  |\phi_\mathrm{res},\mathrm{cl}\rangle
  \frac{e^{-i[\varphi(z,t_{\p{f}})-\varphi(z',t_{\p{i}})]}}{g_\mathrm{stat}(z) i\hbar\dot{E}_\mathrm{res}g_\mathrm{stat}(z')}
\langle\phi_\mathrm{res},\mathrm{cl}|G_\mathrm{stat}(z')\Bigg\}\, . \label{G2Boftexact}
\end{align}
This Green's function represents the exact evolution of the two-body system in the two-channel, single resonance approach. The separable form of the closed-channel potential and the linear time dependence of the resonance energy allow the dynamics to be calculated from the known Green's function $G_\p{stat}(z)$ associated with the stationary Hamiltonian $H_{\p{stat}}$ and its closed-channel component. 

We will now evaluate the transition amplitude $T_\mathrm{diss}(p)$ of Eq.~(\ref{transitionamplitude}) for the case of an asymptotic ramp, which allows the integrals contained in Eq.~(\ref{G2Boftexact}) to be calculated analytically. In the limit of $t_{\p{i}} \to -\infty$, $E_{\p{res}}(t_{\p{i}})$ is large and negative, and so we can write 
\begin{align}
&G_\mathrm{stat}(z)\biggl[1 - |\phi_\mathrm{res},\mathrm{cl}\rangle \frac{1}{g_\mathrm{stat}(z)} \langle\phi_\mathrm{res},\mathrm{cl}|G_\mathrm{stat}(z) \biggr] \nonumber \\
& \underset{t_\p{i} \to -\infty}{\sim} G_\mathrm{stat}(z) \bigg[1 + 
|\phi_\mathrm{res},\mathrm{cl}\rangle
\frac{E_\mathrm{res}(t_{\p{i}})}{1-E_\mathrm{res}(t_{\p{i}})g_\mathrm{stat}(z)}
\langle\phi_\mathrm{res},\mathrm{cl}|G_\mathrm{stat}(z)\bigg] \nonumber \\
&\:\:\: = G_\mathrm{2B}(B_{\p{i}},z)\, . \label{G2Bofz}
\end{align}
The energy-dependent Green's function $G_\mathrm{2B}(B_{\p{i}},z)$ is associated with $H_{\p{2B}}(B_{\p{i}})$. We note that since the singularities of $G_\mathrm{2B}(B,z)$ determine the two-body energy spectrum, the magnetic field dependent energy of the molecular bound state is constrained by the condition
\begin{align}
E_\mathrm{res}(B)g_\mathrm{stat}(E_\mathrm{b}(B))=1 \, ,
\label{boundstate}
\end{align}
which is equivalent to Eq.~(\ref{determinationEb}). The contribution given by Eq.~(\ref{G2Bofz}) to the transition amplitude from the initial bound state $\phi_\p{b}^\p{i}$ to the final continuum state $\phi_\mathbf{p}^\p{f}$ is then 
\begin{align}
i\hbar\int & \frac{dE}{2\pi\hbar} e^{-iE(t_{\p{f}}-t_{\p{i}})/\hbar}
\langle\phi_\mathbf{p}^\p{f}|G_\mathrm{2B}(B_{\p{i}},z)|\phi_\mathrm{b}^\p{i}\rangle \nonumber \\ 
& = e^{-iE_\mathrm{b}(B_{\p{i}})(t_{\p{f}}-t_{\p{i}})/\hbar}\langle\phi_\mathbf{p}^\p{f}|\phi_\mathrm{b}^\p{i}\rangle \, . \label{firstbit}
\end{align}
In the asymptotic limit of $t_{\p{i}}\rightarrow -\infty$ and
$t_{\p{f}}\rightarrow +\infty$,  the interchannel coupling becomes negligible at the initial and final fields. The initial Fesh\-bach molecular state $\phi_\mathrm{b}^\p{i}$ and the final continuum state $\phi_\mathbf{p}^\p{f}$ are then orthogonal, and the contribution of Eq.~(\ref{G2Bofz}) to
$T_\mathrm{diss}(p)$ vanishes.

The remaining term in Eq.~(\ref{G2Boftexact})  contains the rapidly varying phase terms $\varphi(z',t_{\p{i}})$ and $\varphi(z,t_{\p{f}})$, which we evaluate using the stationary phase method~\cite{review}. First we evaluate the matrix elements that appear in the integrals over $E$ and $E'$. The point of stationary phase for the term $\varphi(z',t_\p{i})$ occurs at $z' = E_\p{b}(B_\p{i})$. Using the definition of the Fesh\-bach molecular state in Eq.~(\ref{phib}), and Eqs.~(\ref{twochannelnormalisation}) and (\ref{phase}) we find that
\begin{align}
\frac{\big<\phi_{\p{res}},\p{cl}|G_{\p{stat}}(z')|\phi_{\p{b}}^{\p{i}}\big>}{g_{\p{stat}}(z')} & \underset{z' \to E_{\p{b}}^{\p{i}}}{\sim} \mathcal{N}_{\p{b}}(B_\p{i}) \nonumber \\
& \:\:\:=  \sqrt{-\hbar\dot{E}_{\p{res}}\varphi''(E_{\p{b}}^{\p{i}}, t_{\p{i}})} \, . \label{mat1}
\end{align}
Here we have also made use of the following representation of $G_\p{stat}(z)$ in terms of the entrance-channel Green's function $G_\mathrm{bg}(z)$ and the coupling $W$:
\begin{align}
&G_\mathrm{stat}(z)= |\mathrm{bg}\rangle
  G_\mathrm{bg}(z)\langle\mathrm{bg}|
  +|\phi_\mathrm{res},\mathrm{cl}\rangle\frac{1}{z}\langle\phi_\mathrm{res},\mathrm{cl}|\nonumber\\
&+G_\mathrm{bg}(z)W|\phi_\mathrm{res},\mathrm{bg}\rangle\langle\phi_\mathrm{res},\mathrm{cl}|
+|\phi_\mathrm{res},\mathrm{cl}\rangle\frac{1}{z}\langle\phi_\mathrm{res},\mathrm{bg}|
WG_\mathrm{bg}(z)\nonumber\\
&+|\phi_\mathrm{res},\mathrm{cl}\rangle\frac{1}{z}\langle\phi_\mathrm{res}|WG_\mathrm{bg}(z)
W|\phi_\mathrm{res}\rangle\langle\phi_\mathrm{res},\mathrm{cl}|G_\mathrm{stat}(z)\,.
\label{Gres}
\end{align}
Using the definition of the scattering states in Eq.~(\ref{phip}) we also find, in the asymptotic limit of $t_\p{f} \to +\infty$ where the interchannel coupling can be neglected,
\begin{align}
&\frac{\big<\phi_{\mathbf{p}}^\p{f}|G_{\p{stat}}(z)|\phi_{\p{res}},\p{cl}\big>}{g_{\p{stat}}(z)} \underset{t_{\p{f}} \to +\infty}{\sim} \frac{\big<\phi_{\mathbf{p}}^{(+)}|W|\phi_{\p{res}}\big>}{z - p^2/m} \, . \label{mat2}
\end{align}

The results of Eqs.~(\ref{mat1}) and (\ref{mat2}) then give the transition amplitude to be
\begin{align}
T_\mathrm{diss}(p) = & -\frac{\big<\phi_{\mathbf{p}}^{(+)}|W|\phi_{\p{res}}\big> \mathcal{N}_{\p{b}}(B_\p{i})}{2\pi\hbar\dot{E}_\p{res}} e^{i\varphi(E_\p{b}^\p{i}, t_{\p{i}})} \int dE \frac{e^{-i\varphi(z,t_\p{f})}}{z - p^{2}/m} \times \nonumber \\ 
& \times \int_{-\infty}^{E_\p{b}^\p{i}} dE' e^{-\frac{1}{2} i |\varphi''(E_{\p{b}}^\p{i},t_\p{i})|(E' - E_\p{b}^\p{i})^2} \, . 
\end{align}
Reversing the order of integration, the energy integrals can be evaluated in the limit of $t_\p{i} \to -\infty$, giving
\begin{align}
 T_\mathrm{diss}(p) = &
-\sqrt{\frac{2\pi}{\hbar|\dot{E}_\mathrm{res}|}} \langle\phi_\mathbf{p}^{(+)}|W|\phi_\mathrm{res}\rangle \,\times \nonumber \\ 
    & \times \exp\left(i\big[\varphi\big(E_\mathrm{b}(B_{\p{i}}),t_{\p{i}}\big)-\varphi\big(p^2/m+i0,t_{\p{f}}\big) + \pi/4\big]\right) \label{Tdissfinal}\, .   
\end{align}
The imaginary part of the phase difference will give a real damping factor when we calculate $|T_\mathrm{diss}(p)|^2$, and is therefore the quantity of interest. Using the spectral decomposition of $G_\mathrm{bg}(z)$ to evaluate the imaginary part of the phase factor in Eq.~(\ref{Tdissfinal}) then gives the result of Eq.~(\ref{probabilityexact}), which leads to Eq.~(\ref{finalspectraldensity}) for energies $E$ in the Wigner threshold law regime. We note that the accuracy of Eq.~(\ref{finalspectraldensity}) depends on the quality of the asymptotic evaluation of the phase integral in Eq.~(\ref{Tdissfinal}). For a given resonance and initial and final magnetic field strengths, the associated conditions are easier to fulfil for slower magnetic field ramps.


\section{Asymptotic dissociation spectrum in the single-channel approach}
\label{app:SingleChannel}

The effective single-channel approach aims to recover the
resonance-enhanced scattering properties by choosing the interaction
potential as a perturbed background scattering potential. If such an
interaction potential is taken to have a
separable form, the two-body Hamiltonian can be chosen to be \cite{KGB03,KGG04,SzymanskaPRA05}
\begin{align}
  H_\mathrm{2B}=H_0+V_\mathrm{eff}(B) \, .
  \label{H2B1channel}
\end{align}
Here the associated effective interaction potential is
\begin{align}
  V_\mathrm{eff}(B)=|\chi_\mathrm{bg}\rangle \xi(B)
  \langle\chi_\mathrm{bg}| \, ,
  \label{Veff}
\end{align}
and $H_0=-\hbar^2 \nabla^2/m$ is the kinetic energy operator of the
relative motion of an atom pair. $V_\mathrm{eff}(B)$ is characterised by the form factor $\chi_\mathrm{bg}$ and the amplitude $\xi(B)$. The magnetic-field dependence of $V_\mathrm{eff}(B)$ is contained in $\xi(B)$ through its dependence on the
scattering length $a(B)$~\cite{KGB03}:
\begin{align}
\frac{1}{\xi(B)}=\langle\chi_\mathrm{bg}|G_0(0)|\chi_\mathrm{bg}\rangle
 + \frac{m(2\pi\hbar)^3
  |\langle\chi_\mathrm{bg}|0\rangle|^2}{4\pi\hbar^2 a(B)}\, .
\label{xi}
\end{align} 
Here $G_0(z) = (z - H_0)^{-1}$ is the free Green's function associated with the
Hamiltonian $H_0$ and evaluated at zero energy in Eq.~(\ref{xi}), while
$\langle\chi_\mathrm{bg}|0\rangle$ denotes the overlap of the form
factor $\chi_\mathrm{bg}$ with a zero-momentum plane wave.

Given the resonance-enhanced behaviour of the scattering
length of Eq.~(\ref{eq:a}), Eq.~(\ref{xi}) in
general implies a nonlinear dependence of $\xi(B)$ on
the magnetic field strength $B$. However, if $B$ varies linearly with
time, it can be shown that in the close vicinity of the singularity of the scattering length at $B=B_0$, the amplitude of the separable potential $\xi(B)$ follows the linear dependence given by
\begin{align}
\xi(t)=\xi_{0}+\dot{\xi}(t-t_0) \, .
\label{xioft}
\end{align} 
Here $1/\xi_{0}=\langle\chi_\mathrm{bg}|G_0(0)|\chi_\mathrm{bg}\rangle$,
$t_0$ denotes the time when the zero-energy resonance is crossed, and
\begin{align}
\dot{\xi}=\frac{m\dot{B}(2\pi\hbar)^3
|\langle\chi_\mathrm{bg}|0\rangle|^2}{4\pi\hbar^2 a_\mathrm{bg} \Delta B
|\langle\chi_\mathrm{bg}|G_0(0)|\chi_\mathrm{bg}\rangle|^2}\, .
\label{xidot}
\end{align} 
In the following, we will assume that the linear form of Eq.~(\ref{xioft}) applies throughout the whole dissociation ramp. It can be shown that the final result is independent of the precise time dependence of $\xi(t)$ outside the near-resonant region.

Given the linear time dependence of Eq.~(\ref{xioft}) and the separable form of the interaction potential in Eq.~(\ref{Veff}), the effective single-channel Hamiltonian is similar to the two-channel Hamiltonian of Eqs.~(\ref{H2Boft}) and (\ref{eq:hcl}). Consequently, the analytic determination of the evolution operator for the single-channel Hamiltonian of Eq.~(\ref{H2B1channel}) proceeds along the lines of Appendix \ref{app:EvolutionOperator} by noting the following substitutions: $H_0$ plays the role of $H_\mathrm{stat}$, while $E_\mathrm{res}(t)$ is replaced by $\xi(t)$ and $|\phi_\mathrm{res},\mathrm{cl}\rangle$ by $|\chi_\mathrm{bg}\rangle$. A similar substitution is applied to the respective Green's functions and their matrix elements.

We will now evaluate the general expression of Eq.~(\ref{transitionamplitude}) for the dissociation transition amplitude in the asymptotic limit of $t_{\p{i}}\rightarrow -\infty$ and $t_{\p{f}}\rightarrow +\infty$ for the single-channel Hamiltonian. Applying the above substitutions, the contribution of Eq.~(\ref{G2Bofz}) to $T_\mathrm{diss}(p)$ vanishes for the same reasons as in the two-channel case. The first phase integral of the remaining term in Eq.~(\ref{G2Bofz}) is analogous to the two-channel case, and is given by Eq.~(\ref{mat1}) with the appropriate substitution. The second phase integral (corresponding to Eq.~(\ref{mat2}) in the two-channel case) takes the form
\begin{align}
\int_{E_\mathrm{b}(B_{\p{i}})}^{\infty}&\frac{dE}{2\pi\hbar}\, e^{-i \varphi(z,t_{\p{f}})}
\frac{\langle\phi_\mathbf{p}^{(+)}|G_0(z)|\chi_\mathrm{bg}\rangle}{g_0(z)}\nonumber\\ 
&\underset{t_{\p{f}} \to \infty}{\sim}
-\frac{i}{\hbar} e^{-i
  \varphi(p^2/m+i0,t_{\p{f}})}
\frac{\langle\mathbf{p}| \chi_\mathrm{bg}\rangle}{g_0(p^2/m+i0)} \, .
\end{align}
This leads to the following result for $T_\mathrm{diss}(p)$:
\begin{align}
T_\mathrm{diss}(p) &=
-\sqrt{\frac{2\pi}{\hbar|\dot{\xi}|}}\frac{\langle\mathbf{p}| \chi_\mathrm{bg}\rangle}{g_0(p^2/m+i0)} \, \times  \nonumber \\ 
& \times \exp\left(i\big[\varphi(E_\mathrm{b}(B_{\p{i}}),t_{\p{i}})-\varphi(p^2/m+i0,t_{\p{f}})+\pi/4\big]\right) \label{Tdiss} \, .
\end{align}

To calculate the imaginary part of the phase difference we use the definition of $\varphi(E, t)$ of Eq.~(\ref{phase}), yielding
\begin{align}
\mathrm{Im}\big[\varphi(E_\mathrm{b}(B_{\p{i}}),t_{\p{i}}) &-\varphi(p^2/m+i0,t_{\p{f}})\big]  \nonumber \\
&=\frac{1}{\hbar\dot{\xi}}\mathrm{Im}\int_0^{p^2/m}
\frac{dE}{g_0(E+i0)} \, .
\end{align}
Multiplying the numerator and denominator of the integrand by $-\xi(t_{\p{f}})$, we then take the asymptotic limit of $t_{\p{f}} \rightarrow +\infty$, in which case $\xi(t_{\p{f}})$ is large and we can write
\begin{align}
\mathrm{Im}\big[\varphi&(E_\mathrm{b}(B_{\p{i}}),t_{\p{i}})-\varphi(p^2/m+i0,t_{\p{f}})\big]  \nonumber \\
&\underset{t_\p{f} \to \infty}{\sim} \frac{1}{\hbar\dot{\xi}g_0(0)}\mathrm{Im}\int_0^{p^2/m} dE
\frac{1 - \xi(t_{\p{f}})g_0(0)}{1 - \xi(t_{\p{f}})g_0(E+i0)} \, .
\end{align}
We then make use of the general form of the $T$-matrix of any separable potential~\cite{GKGTJ04},
\begin{align}
\left<0\left|T_{\p{eff}}(E+i0)\right|0\right> &= \frac{\xi(t_{\p{f}})|\langle\chi_{\p{bg}}|0\rangle|^{2}}{1 - \xi(t_{\p{f}})g_0(E+i0)} \, .
\end{align}
Together with the explicit expression for $\dot{\xi}$ given in Eq.~(\ref{xidot}), and the low-energy expansion of the $T$-matrix, this gives the following result in the Wigner threshold law regime:
\begin{align}
&\mathrm{Im}\left[\varphi(E_\mathrm{b}(B_{\p{i}}),t_{\p{i}})-\varphi(p^2/m+i0,t_{\p{f}})\right]=
\frac{2}{3}\frac{a_\mathrm{bg} \Delta B}{\hbar^2 m \dot{B}} p^3 \, .
 \label{eq:lowE}
  \end{align}
By evaluating the remaining prefactors in Eq.~(\ref{Tdiss}) in the zero-momentum limit, we arrive at the final formula for the dissociation transition probability density,
\begin{align}
|T_\mathrm{diss}(p)|^2=\frac{|a_\mathrm{bg} \Delta B|}{\pi
 \hbar^2 m |\dot{B}|} \exp\left(-\frac{4}{3}\frac{a_\mathrm{bg} \Delta
 B}{\hbar^2 m \dot{B}} p^3\right) \, ,  
  \end{align}
which recovers Eq.~(\ref{finalspectraldensity}).

\end{appendix}


\begin{thebibliography}{99}
\bibitem{review}
T. K\"ohler, K. G\'oral, and P.~S. Julienne, e-print cond-mat/0601420 (to be published in Rev. Mod. Phys.).

\bibitem{doyle}
J.~Doyle, B.~Friedrich, R.~V.~Krems, and F.~Masnou-Seeuws, Eur. Phys. J. D \textbf{31}, 149 (2004).

\bibitem{regal03}
C.~A.~Regal, C.~Ticknor, J.~L.~Bohn, and D.~S.~Jin, Nature (London) \textbf{424}, 47 (2003).

\bibitem{gre03}
M.~Greiner, C.~A.~Regal, and D.~S.~Jin, Nature (London), \textbf{426}, 537 (2003).

\bibitem{cub03}
J.~Cubizolles, T.~Bourdel, S.~J.~J.~M.~F. Kokkelmans, G.~V. Shlyapnikov, and C.~Salomon, Phys. Rev. Lett. \textbf{91}, 240401 (2003).

\bibitem{xu03}
K.~Xu, T.~Mukaiyama, J.~R.~Abo-Shaeer, J.~K.~Chin, D.~E.~Miller, and W.~Ketterle, Phys. Rev. Lett. \textbf{91}, 210402 (2003).

\bibitem{str03}
K.~E.~Strecker, G.~B.~Partridge, and R.~G.~Hulet, Phys.~Rev.~Lett. \textbf{91}, 080406 (2003).

\bibitem{herbig03}
J.~Herbig, T.~Kraemer, M.~Mark, T.~Weber, C.~Chin, \mbox{H.-C.}~N\"agerl, and R.~Grimm, Science \textbf{301}, 1510 (2003).

\bibitem{duerr04prl}
S.~D\"urr, T.~Volz, A.~Marte, and G.~Rempe, Phys.~Rev.~Lett.\textbf{92}, 020406 (2004).

\bibitem{MukaiyamaPRL04}
T.~Mukaiyama, J.~R.~Abo-Shaeer, K.~Xu, J.~K. Chin, and W.~Ketterle,
Phys. Rev. Lett. \textbf{92}, 180402 (2004).

\bibitem{DuerrPRA04}
S. D{\"u}rr, T. Volz, and G. Rempe, Phys. Rev. A \textbf{70}, 031601
(2004).

\bibitem{volz05}
  T.~Volz, S.~D\"{u}rr, N.~Syassen, G.~Rempe, E.~van~Kempen, and S.~Kokkelmans, Phys.~Rev.~A \textbf{72}, 010704 (2005).

\bibitem{donley02}
E.~A.~Donley, N.~R.~Claussen, S.~T.~Thompson, and C.~E.~Wieman, Nature (London) \textbf{417}, 529 (2002).

\bibitem{chin03}
C.~Chin, A.~J.~Kerman, V.~Vuleti\'c, and S.~Chu, Phys.~Rev.~Lett. \textbf{90}, 033201 (2003).

\bibitem{jochim03}
S.~Jochim, M.~Bartenstein, A.~Altmeyer, G.~Hendl, C.~Chin, J.~Hecker~Denschlag, and R.~Grimm, Phys.~Rev.~Lett \textbf{91}, 240402 (2003).

\bibitem{jochim03b}
S.~Jochim, M.~Bartenstein, A.~Altmeyer, G.~Hendl, S.~Riedel, C.~Chin, J.~Hecker~Denschlag, and R.~Grimm, Science \textbf{302}, 2101 (2003).

\bibitem{mark05}
M.~Mark, T.~Kraemer, J.~Herbig, C.~Chin, H.-C. N\"agerl, and R.~Grimm, Europhys. Lett. \textbf{69}, 706 (2005).

\bibitem{thompson05}
S.~T.~Thompson, E.~Hodby, and C.~E.~Wieman, Phys.~Rev.~Lett. \textbf{95}, 190404 (2005).

\bibitem{duerr05partial}
  S.~D\"{u}rr, T.~Volz, N.~Syassen, G.~Rempe, E.~van~Kempen, S.~Kokkelmans, B.~Verhaar, and H.~Friedrich, Phys.~Rev.~A \textbf{72}, 052707 (2005).

\bibitem{stoll05}
M.~Stoll and T.~K\"ohler, Phys.~Rev.~A \textbf{72}, 022714 (2005).

\bibitem{thomas}
L. H. Thomas, Phys. Rev. \textbf{47}, 903 (1935).

\bibitem{efimov}
V. Efimov, Phys. Lett. \textbf{33B}, 563 (1970).

\bibitem{koehler03}
T.~K\"ohler, T.~Gasenzer, P.~S.~Julienne, and K.~Burnett, Phys. Rev. Lett. \textbf{91}, 230401 (2003).

\bibitem{kraemer06}
T.~Kraemer, M.~Mark, P.~Waldburger, J.~G.~Danzl, C.~Chin, B.~Engeser, A.~D.~Lange, K.~Pilch, A.~Jaakkola, H.-C.~N\"agerl, and R.~Grimm,  Nature (London) \textbf{440}, 315 (2006).

\bibitem{smirne06}
G.~Smirne, R.~M.~Godun, D.~Cassettari, V.~Boyer, C.~J.~Foot, T.~Volz,
N.~Syassen, S.~D\"urr, G.~Rempe, M.~D.~Lee, K.~G\'oral, and T.~K\"ohler, e-print cond-mat/0604183.

\bibitem{GKGTJ04}
  K. G\'oral, T. K\"ohler, S.~A. Gardiner, E. Tiesinga, and P.~S. Julienne,
  J. Phys. B \textbf{37}, 3457 (2004).

  \bibitem{haq05}
M. Haque and H.~T.~C. Stoof, Phys.~Rev.~A \textbf{71}, 063603 (2005).

\bibitem{brouard05}
  S.~Brouard and J.~Plata, Phys.~Rev.~A \textbf{72}, 023620 (2005).

\bibitem{WignerPR48}
E. P. Wigner, Phys. Rev. \textbf{73}, 1002 (1948).

\bibitem{jul89}
P.~S.~Julienne and F.~H.~Mies, J.~Opt.~Soc.~Am.~B \textbf{6}, 2257 (1989).

\bibitem{gri93}
G.~F.~Gribakin and V.~V.~Flambaum, Phys.~Rev.~A \textbf{48}, 546 (1993).

\bibitem{KGG04}
  T.~K\"ohler, K.~G\'oral, and T.~Gasenzer, Phys. Rev. A \textbf{70}, 023613 (2004).
  
\bibitem{petrov04}
D.~S.~Petrov, Phys.~Rev.~Lett. \textbf{93}, 143201 (2004).
  
\bibitem{Duerrprivate}
S. D{\"u}rr, private communication (2006).

\bibitem{dem68}
Y. N. Demkov and V. I. Osherov, Sov. Phys. -JETP \textbf{26}, 916 (1968).

\bibitem{mac98}
J.~H.~Macek and M.~J.~Cavagnero, Phys. Rev. A \textbf{58}, 348 (1998).

\bibitem{KGB03}
  T.~K\"ohler, T.~Gasenzer, and K.~Burnett, Phys.~Rev.~A \textbf{67}, 013601 (2003).

\bibitem{SzymanskaPRA05}
M.~H. Szyma{\'n}ska, K. G\'{o}ral, T. K\"{o}hler, and K. Burnett,
Phys.~Rev.~A {\bf 72}, 013610 (2005).

\end{thebibliography}
\end{document}